\documentclass[letterpaper, 10 pt, conference]{ieeeconf}  % Comment this line out if you need a4paper

\IEEEoverridecommandlockouts                              % This command is only needed if 
\title{\LARGE \bf
Generalizing Surgical Instruments Segmentation to Unseen Domains with One-to-Many Synthesis
}

\author{An Wang*$^{1}$, Mobarakol Islam*$^{2}$, Mengya Xu$^{3}$, and Hongliang Ren$^\dagger$$^{1,3}$% <-this % stops a space
\thanks{This work was supported by the Shun Hing Institute of Advanced Engineering (SHIAE project BME-p1-21) at the Chinese University of Hong Kong (CUHK), Hong Kong Research Grants Council (RGC) Collaborative Research Fund (CRF C4026-21GF and CRF C4063-18G), (GRS) \#3110167 and Shenzhen-Hong Kong-Macau Technology Research Programme (STIC Type C 202108233000303, SGDX20210823103535014).}% <-this % stops a space
\thanks{* These authors contributed equally to this work.}
\thanks{$^\dagger$ Corresponding author: Hongliang Ren (hlren@ieee.org).}
\thanks{$^{1}$Department of Electronic Engineering, Shun Hing Institute of Advanced Engineering (SHIAE), CUHK, Hong Kong, China.
        {\tt\small wa09@link.cuhk.edu.hk, hlren@ee.cuhk.edu.hk}}%
\thanks{$^{2}$Department of Computer Science, Wellcome / EPSRC Centre for Interventional and Surgical Sciences (WEISS), UCL. London, UK.
        {\tt\small mobarakol.islam@ucl.ac.uk}}%
\thanks{$^{3}$Department. of Biomedical Engineering, NUS, Singapore
        {\tt\small mengya@u.nus.edu}}%
}

\usepackage{adjustbox}
\usepackage{multirow}
\usepackage{bbold}
\usepackage{siunitx}

\usepackage{pifont}
\usepackage{colortbl}
\usepackage{hhline}
\usepackage{float}
\usepackage{diagbox}

\usepackage[normalem]{ulem}
\usepackage{color}
\usepackage{booktabs}
\usepackage{bbm}
\usepackage{makecell}
\usepackage{rotating}
\usepackage{hyperref}
\begin{document}

\maketitle
\thispagestyle{empty}
\pagestyle{empty}

\begin{abstract}
Despite their impressive performance in various surgical scene understanding tasks, deep learning-based methods are frequently hindered from deploying to real-world surgical applications for various causes. Particularly, data collection, annotation, and domain shift in-between sites and patients are the most common obstacles. In this work, we mitigate data-related issues by efficiently leveraging minimal source images to generate synthetic surgical instrument segmentation datasets and achieve outstanding generalization performance on unseen real domains. Specifically, in our framework, only one background tissue image and at most three images of each foreground instrument are taken as the seed images. These source images are extensively transformed and employed to build up the foreground and background image pools, from which randomly sampled tissue and instrument images are composed with multiple blending techniques to generate new surgical scene images. Besides, we introduce hybrid training-time augmentations to diversify the training data further. 
Extensive evaluation on three real-world datasets, i.e., Endo2017, Endo2018, and RoboTool, demonstrates that our one-to-many synthetic surgical instruments datasets generation and segmentation framework can achieve encouraging performance compared with training with real data. Notably, on the RoboTool dataset, where a more significant domain gap exists, our framework shows its superiority of generalization by a considerable margin. We expect that our inspiring results will attract research attention to improving model generalization with data synthesizing.

\end{abstract}

% \keywords{Surgical Instrument Segmentation, One-to-Many Synthesis, Image Blending, Data Augmentation, Generalization}

\maketitle

\section{Introduction}
\label{sec:intro}
% Two issues to be handled with synthetic dataset in this work: data shortage and domain shift.
Deep learning models trained with sufficient real-world data have achieved tremendous success in various computer-assisted applications of surgical scene understanding, including instrument segmentation~\cite{islam2021st,jin2022exploring,islam2020learning},
% workflow recognition~\cite{jin2021temporal,ding2022exploring}, 
% report generation~\cite{xu2021learning,lin2022sgt,xu2021class},
image captioning~\cite{xu2022rethinking}, report generation~\cite{xu2021learning,xu2021class}, and visual question answering~\cite{seenivasan2022surgical}. However, demanding challenges, such as the lack of well-annotated data and data shifts, have greatly hindered their practical deployment. One of the major reasons is that most of these works heavily depend on the availability of adequate and well-labeled training data to perform Supervised Learning. In contrast, surgical data collection and annotation are usually time-consuming and labor-intensive. This has decreased the training feasibility and efficiency. Besides, model performance is frequently observed to degrade during deployment to real-world scenarios due to various domain shifts,
% ~\cite{guan2021domain,stacke2020measuring} 
such as intensity shift, acquisition shift, and population shift. Thus, model generalization ability has gained increasing significance during model evaluation.

% Other methods to solve these issues - transfer learning, domain adaptation (model based)
Over the years, researchers have investigated several solutions, i.e., self-supervised transfer learning~\cite{neimark2021train}, class-incremental domain adaptation~\cite{xu2021class}, 
and domain generalization~\cite{philipp2022dynamic}, to narrow the performance discrepancy between developing and deploying deep learning-based surgical applications.
% Among them,  are  the hottest research fields.
% to improve the model performance across multiple domain data. 
% With self-supervised pre-training and the Time-Series Adaptation Network (TSAN), Neimark et al.~\cite{neimark2021train} transfer the model trained with one type of laparoscopic surgery to other procedure types and achieve over 90\% accuracy in recognizing surgical steps. Besides, domain adaptation and generalization techniques are also frequently used to preserve the performance on the target domains with various data shifts. 
For example, to handle novel instruments when generating the surgical report in the new domain, Xu et al.~\cite{xu2021class} propose the class-incremental domain adaptation by incorporating curriculum by smoothing in the transformer-based caption model and adopting the label smoothing to better calibrate the model. However, one of the drawbacks of these works is that they still depend on a sufficient amount of densely annotated real data for model training, limiting their application in the case of data shortage. 
% The above two main streams of works mainly focus on 
% Contrary to focusing on model enhancement regarding better transferable and generalization ability,

To alleviate data-related issues like data shortage and shifts, recently, 
the emerging trend of data-centric AI~\cite{zhang2022shifting}
% ~\cite{polyzotis2021can,eyuboglu2022dcbench,zhang2022shifting} 
has gained increasing attention and inspired another research line, i.e., learning with synthetic data~\cite{colleoni2020synthetic,colleoni2022ssis}. 
% ~\cite{chen2021synthetic,colleoni2020synthetic}. 
For example, SSIS-Seg~\cite{colleoni2022ssis} proposes the simulation-supervised loss and the attention similarity loss in the image-to-image (I2I) translation process to generate high-quality synthetic surgical images. However, like most simulation-to-real settings, this work also needs the real-world dataset as the target domain for style transfer, decreasing its feasibility for unseen domains. 

% In this work, we extend our previous framework~\cite{wang2022rethinking} by exploring three image blending modes and three training-time augmentations.o
% The encouraging performance across multiple real-world datasets demonstrates that our upgraded pipeline is reliable and robust for synthesis-based surgical instruments segmentation.

\begin{figure*}[!htbp]
\centering
\includegraphics[width=\textwidth]{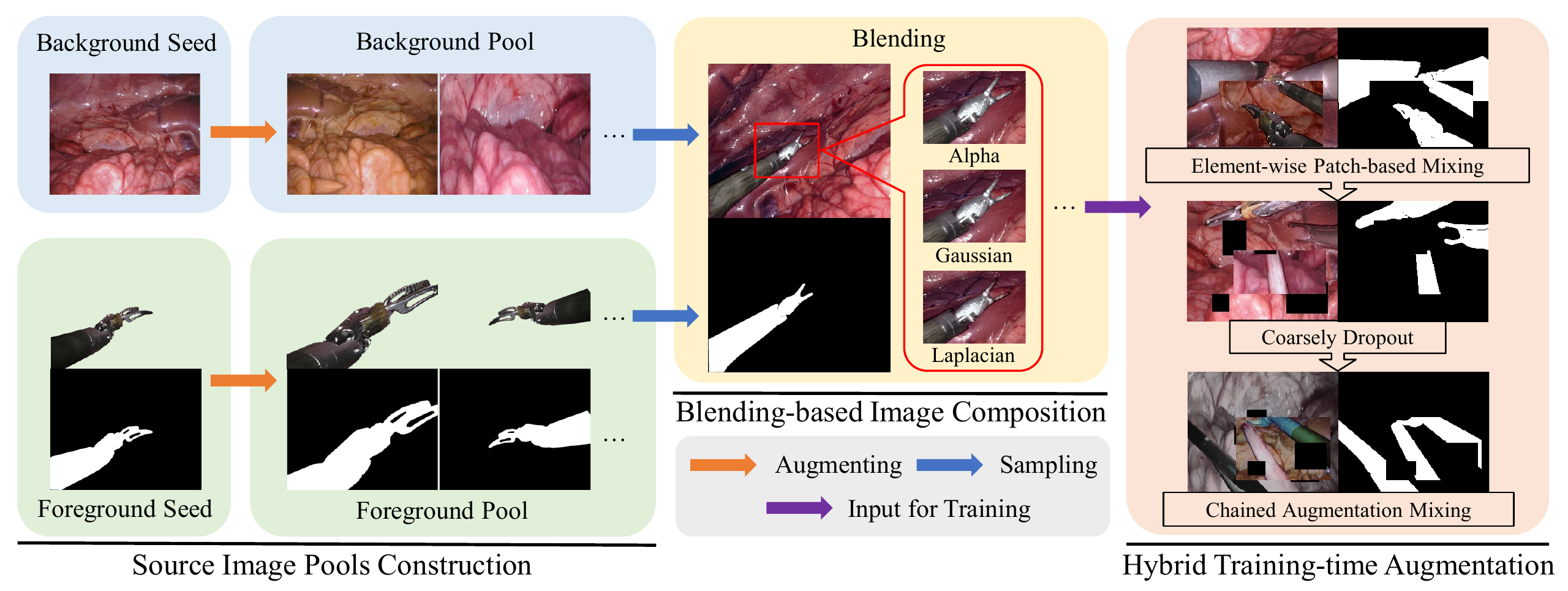}
\caption{\textbf{Overview of the proposed one-to-many surgical instruments synthesis and segmentation framework.} The framework mainly consists of three consecutive steps, i.e., Source Image Pools Construction, Blending-based Image Composition, and Hybrid Training-time Augmentation. 
First, we extensively augment the foreground and background seed images and construct the image pools. Then Alpha, Gaussian, or Laplacian Blending is adopted to compose the randomly sampled images from the two image pools. Finally, the composited images are further transformed with hybrid in-training augmentation to boost their diversity, including Element-wise Patch-based Mixing, Coarsely Dropout, and Chained Augmentation Mixing.
}
\label{fig:proposal}
\end{figure*}

In this paper, to eliminate the dependence on abundant real data and boost generalization performance across multiple domains, we investigate extreme source constraints and set up a highly data-efficient one-to-many surgical instrument synthesis and segmentation framework, as shown in Fig.~\ref{fig:proposal}. Our holistic framework considers improving data quality from two aspects, i.e., the realism of each image during synthesizing and the diversity of the entire dataset during training. 
With only one background tissue image and a few foreground instrument images, our data-centric pipeline consists of three main steps, i.e., Source Image Pools Construction, Blending-based Image Composition, and Hybrid Training-time Augmentation. 
% dataset imbalance / novel instruments / domain shift
% Compared with I2I-based works, our pipeline gets rid of adversarial training in the simulation-to-real translation. 
% The main contributions of this work can be summarized as below:
We can summarize the main contributions of this work below:
% 1. propose a synthesis training framework 
% 2. explore various blending; in-training augmentation and public the code; 
% 3. Summary: single image data-efficient, stability (low STD), generalization
\begin{enumerate}
    \item With extremely limited source images and without costly manual annotation, we propose a one-to-many surgical instruments synthesis and segmentation framework that can achieve promising generalization performance on unseen domains.
    \item We explore three blending techniques to improve the quality of the composited images and find that Laplacian Blending yields the best results in most cases, compared with Alpha Blending and Gaussian Blending. 
    \item We incorporate Chained Augmentation Mixing, Coarsely Dropout, and Element-wise Patch Mixing into a hybrid pipeline to alleviate overfitting and boost training data diversity. The implementation is released and can be easily integrated into other surgical scene understanding tasks. 
    \item Extensive experiments across multiple domains with incremental classes and data shifts suggest that our framework can achieve close and even superior accuracy compared with training with the real dataset. Moreover, in the semi-supervised synthetic-real joint training scheme, we find that a small portion of real data can efficiently boost the model performance by a large margin.  
\end{enumerate}

\section{Related Works}
Several image composition-based synthetic dataset generation frameworks have been proposed in recent years to overcome data restrictions for training deep learning models and demonstrate their simplicity and efficiency. 
The pioneering work~\cite{dwibedi2017cut} presents a ``Cut, Paste and Learn" framework to generate large amounts of annotated data for instance detection, and achieves promising results at a low cost. 
The successive work~\cite{tripathi2019learning} proposes to jointly train the image synthesizer network and the target task network in an adversarial style. And a discriminator network is added to improve the realism of the synthesized images.
In computer-aided surgery, Garcia-Peraza-Herrera et al.~\cite{garcia2021image} propose to compose broadly collected tissue images with manually recorded instrument images to create synthetic surgical scene images with ``mix-blend" to relieve the blending artifacts. 
The above works mitigate the deficiency of annotated data by image composition-based synthesizing. However, one limitation is that they still rely on sufficient source images of foreground and background instances. 
On the contrary, Wang et al.~\cite{wang2022rethinking} deal with a more challenging condition where only one background tissue image and a few foreground instrument images are available. However, its naive blending could introduce artifacts and cause over-fitting. 
% The model trained with the synthetic surgical dataset shows decent segmentation performance on two real datasets. 

Training-time data augmentation is essential in boosting model performance regarding in-domain accuracy, out-of-domain generalization, and robustness. Considering which images to operate on, there are generally two strategies for augmenting data during training, i.e., intra-augmentation, which happens within one specific training sample, and inter-augmentation, which applies to two different training samples. With the former strategy, Cutout~\cite{devries2017improved} random drops square patches in the training image as augmentation. Augmix~\cite{hendrycks2019augmix} proposes to feed the training sample into a couple of augmentation chains, and the outputs of the chains are mixed with the original input to form the final augmented image. With the latter strategy, Cutmix~\cite{yun2019cutmix} cuts cross-image patches and then exchanges them to create new training samples.
Based on these works and considering the typical cases in real surgical scenes like instruments occlusion and overlapping, we design a hybrid integration of them to improve the surgical instrument segmentation performance.

\section{Methods}

\subsection{One-to-Many Dataset Synthesizing}
Our synthetic surgical dataset generation is particularly data-efficient in that only one tissue image and a few instrument images are adopted as the sources. Besides, we don't need to annotate the generated images manually. All the segmentation masks are automatically derived along with the generation of the corresponding synthetic images. 

\subsubsection{Source Image Pools Construction}
Like most image composition-based approaches, the first step of our framework is to prepare the source images for blending.
One distinct advantage of our pipeline is that we start from a limited number of foreground and background images as the seeds, as shown in Fig.~\ref{fig:proposal}.
% to construct the much larger source image pools with a wide variety of image augmentations. 
The background seed image is a pure tissue image without the presence of any instruments. Since it is the only background tissue source image, we purposely choose it to contain diverse tissue appearances and structures. The foreground instrument seed images are pure instrument images with transparent background and their corresponding masks are also readily acquired simultaneously. 
To make them as representable for more instrument states as possible, especially, we consider practical surgical scene properties when choosing these seed images. For instance, the clasper part of some instruments includes two states, i.e., open and closed states. In addition, macro and micro views of the instruments are also considered to cover detailed appearances and different viewpoints. 

% Data augmentation is a common and convenient approach to generate abundant data variations while boosting in-distribution data diversity and out-of-distribution model generalization. 
% Besides, class imbalance, which is frequently seen in medical and surgical datasets, can also be effectively handled by purposely augmenting the minority classes. 
To construct our source image pools, we employ various image intensity and geometry augmentations from the imgaug~\footnote{https://github.com/aleju/imgaug} library. Specifically, we randomly apply various strong augmentations for the single background tissue image, including \textit{HorizontalFlipping}, \textit{VerticalFlipping}, \textit{Cropping}, \textit{AddToHueAndSaturation}, \textit{LinearContrast}, \textit{PerspectiveTransform}, \textit{Affine}, \textit{GaussianBlur}, \textit{AverageBlur}, \textit{MedianBlur}, \textit{Sharpen}, \textit{Emboss}, and \textit{AdditiveGaussianNoise}, to name a few. 
With this, the background seed image $x_b$ is transformed into a background image pool with $m$ augmented images, i.e., $X_b^m = \{x_b^1, x_b^2, x_b^3, ..., x_b^m\}$. 
Similarly, for each foreground instrument seed images $x_f$, we also randomly apply augmentations like \textit{Flipping}, \textit{Cropping}, \textit{Blurring}, \textit{AffineTransformation}, etc. Note that the same geometric transformations are applied to both the instrument images and their segmentation masks. In this way, without additional effort, accurate annotations are easily acquired. Finally, we set up the foreground image pool containing $n$ augmented variants of the instruments with corresponding masks, i.e., $X_f^n = \{x_f^1, x_f^2, x_f^3, ..., x_f^n\}$ and the masks $Y_f^n = \{y_f^1, y_f^2, y_f^3, ..., y_f^n\}$.

\subsubsection{Blending-based Image Composition}
With the constructed background and foreground image pools, i.e., $X_f^n$ and $X_b^m$, we then randomly sample tissue and instrument images from these pools and compose them to create $k$ synthetic surgical scene images $X_{syn}^k$ and corresponding masks $Y_{syn}^k$. 
Image composition by blending is widely known and used in computer graphics and image processing. 
However, when fed into the deep neural networks for training, the blended surgical scene images show insufficiency due to the lack of realism, especially at the instrument contours where blending artifacts exist. As shown in Fig.~\ref{fig:proposal}, we investigate three blending modes, i.e., Alpha Blending, Gaussian Blending, and Laplacian Blending to find the optimal choice for our surgical image composition.
% For simplicity, the detailed description of these conventional blending techniques will be provided in the supplementary material. 

Specifically, \textit{Alpha Blending} happens in the alpha channel of the foreground image $x_f^i \in X_f^n$ and background image $x_b^j \in X_b^m$. The instrument area in the foreground mask $y_f^i \in Y_f^n$ will be directly taken into the final composited image $x_{syn}^t \in X_{syn}^k$, and the remaining area is inherited from the background tissue. In \textit{Gaussian Blending}, the foreground instrument mask $y_f^i$ is eroded and blurred to generate a new mask $\overline{y}_f^i$, with a kernel size of 3 and 5, respectively. Then the foreground image $x_f^i$ and background image $x_b^j$ are combined following Alpha Blending with the transformed mask $\overline{y}_f^i$. Regarding \textit{Laplacian Blending}, we first build the Laplacian pyramid for both the foreground image $x_f^i$ and the background image $x_b^j$. Then we construct a Gaussian pyramid for the area covered by the foreground mask $y_f^i$ and form a combined pyramid with the nodes of the Gaussian pyramid as weights. Lastly, we collapse the combined pyramid to get the final blended image $x_{syn}^t$. 

\begin{figure*}[!htbp]
\centering
\includegraphics[width=\textwidth]{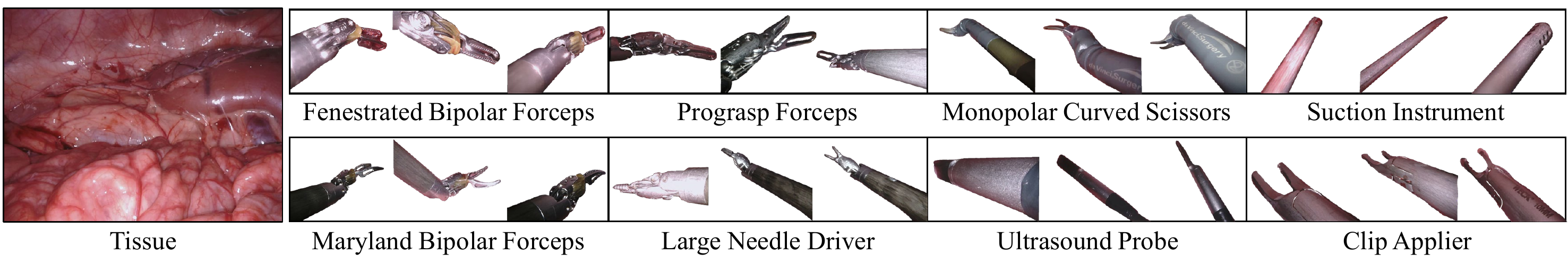}
\caption{\textbf{Seed images of the tissue and eight types of surgical instruments.} Only one single background tissue image and at most three foreground instrument images are utilized as the seed images in our synthetic dataset generation pipeline.}
\label{fig:src_imgs}
\end{figure*}

\subsection{Hybrid Training-time Synthetic Data Augmentation}
With the generated surgical scene images $X_{syn}^k$ and their instrument segmentation masks $Y_{syn}^k$, we feed them into the neural network to train the segmentation model. To improve the generalization ability from the synthetic domain to multiple real domains, we investigate three advanced training-time augmentations and introduce their hybrid integration for training surgical instrument segmentation models. 

\textit{Intra-augmentation} is the most frequently used technique for augmenting training samples. We explore two advanced training-time intra-augmentations. 
Inspired by Cutout~\cite{devries2017improved}, we first introduce the patch-level Coarsely Dropout (CDO) of training samples as a regularization method to alleviate overfitting. This transformation can simulate the practical scenarios of instrument occlusion. In our implementation, the number, shape, and size of the patches to be dropped are all randomly set and applied simultaneously to images and masks during training. 
Besides, we also apply the Chained Augmentation Mixing (CAM), which was initially proposed in Augmix~\cite{hendrycks2019augmix} for image classification. Their label-mixing strategy is not applicable to our segmentation task. Hence, we intentionally tailor their augmentation list to preserve mask accuracy. Specifically,
% In their work, besides mixing images variants, they also mix the image labels for their classification task. Whereas, for our segmentation task, directly mixing masks can cause mask inaccuracy due to random geometric changes in the augmentation chains. Since we have included geometric transformations like flipping and cropping when constructing the source image pools, for simplicity, we adapt their implementation by removing those transformations. 
the transformations in our three augmentation chains include \textit{Autocontrast}, \textit{Equalize}, \textit{Posterize}, and \textit{Solarize} for a soft version and additional \textit{Color}, \textit{Contrast}, \textit{Brightness}, and \textit{Sharpness} for a hard version. 
The outputs of these augmentation chains are mixed together with the original image to form the final training samples. 

Besides intra-augmentation, 
% inter-augmentation also helps improve the model performance. Hence, 
we also adopt one \textit{inter-augmentation}, i.e., the Element-wise Patch Mixing (EPM), originated from Cutmix~\cite{yun2019cutmix}, as another training-time augmentation to simulate instrument overlapping which exists in most practical surgical scenes. 
% , 
Batch-wise image patch mixing has been used in many classification tasks, showing consistent performance gain. We introduce it into our surgical instrument segmentation task with two significant adaptations. Firstly, 
different from the classification task, 
for our segmentation task, we need to manipulate the masks with the same transformations as the corresponding images. Moreover, rather than batch-wise operation, we use a more flexible element-wise operation where patch cutting and mixing are randomly applied to each image in the batch. 

As shown in Fig.~\ref{fig:proposal}, we integrate all these three training-time augmentations in our framework to boost the training data diversity and expand the inherent representation space. With their hybrid usage, we observe steady and increasing performance gains on multiple real-world datasets.

\section{Experiments}
% To demonstrate the efficacy of our proposed framework for one-to-many surgical instruments dataset generation and segmentation, we conduct extensive experiments to compare the segmentation performance of the models trained with our synthetic datasets and the real datasets. 
% The evaluation results across multiple domains indicate that training segmentation models with our synthetic datasets can yield decent and promising performance in a highly efficient manner.

\subsection{Datasets}

\subsubsection{Synthetic Datasets} The foreground and background seed images are effortlessly extracted from the Endo18~\cite{allan20202018} training dataset and contain all eight types of instruments, i.e., \textit{Fenestrated Bipolar Forceps}, \textit{Maryland Bipolar Forceps}, \textit{Prograsp Forceps}, \textit{Large Needle Driver}, \textit{Monopolar Curved Scissors}, \textit{Ultrasound Probe}, \textit{Suction Instrument}, and \textit{Clip Applier}, as shown in Fig.~\ref{fig:src_imgs}.
% By applying various transformations covering intensities and geometries on these source images, we build up two synthetic source image pools, i.e., the foreground instrument image pool and the background tissue image pool. During data generation, one background image and one or two foreground images are randomly sampled from these image pools and blended together to create the synthetic image. Three blending techniques, i.e., Alpha Blending, Gaussian Blending, and Laplacian Blending are utilized in the blending process. 
In natural surgical scenes, usually more than one instrument coexists in the view. Considering this, in our synthetic surgical images, one or two different foreground instruments are blended with each background image. Overall, we construct five synthetic datasets under different numbers of seed and foreground image settings, as summarized in Table~\ref{tab:ds_syn}. In the case of two seed images per instrument, Syn-S2-F1 and Syn-S2-F2 are synthesized by blending one or two unique foreground instruments on the background tissue image. Similarly, when three seed images are adopted per instrument, each synthetic image in Syn-S3-F1 and Syn-S3-F2 contains one or two different foreground instruments. Lastly, Syn-S3-F1F2 is a mixed dataset where 20\% synthetic images have one instrument, and the rest 80\% include two varied instruments in each image. The ratio is empirically set based on the observation that in most surgical scenes, there exists more than one instrument. While our framework can easily generate abundant data, for a fair comparison with real data, we keep the training sample size of the synthetic datasets consistent with the training split of Endo18~\cite{allan20202018}, which is 2235.

% Table generated by Excel2LaTeX from sheet 'ds_info'
\begin{table}[!t]
  \centering
  \caption{\textbf{Detailed settings of our five synthetic datasets.} Only one tissue image is selected as the background seed.}
      \resizebox{0.85\columnwidth}{!}{%
    \begin{tabular}{ccc}
    \toprule
    Synthetic dataset & \begin{tabular}[c]{@{}c@{}}\# seed images \\ per instrument\end{tabular} & \begin{tabular}[c]{@{}c@{}}\# foreground tools \\ per synthetic image\end{tabular}\\
    \midrule
    Syn-S2-F1  & 2     & \multicolumn{1}{c}{1} \\
    Syn-S2-F2  & 2     & \multicolumn{1}{c}{2} \\
    \midrule
    Syn-S3-F1 & 3     & \multicolumn{1}{c}{1} \\
    Syn-S3-F2 & 3     & \multicolumn{1}{c}{2} \\
    \midrule
    Syn-S3-F1F2  & 3     & 1 (20\%) and 2 (80\%) \\%\begin{tabular}[c]{@{}c@{}}20\% 1 \\ + 80\% 2 \end{tabular}\\
    \bottomrule
    \end{tabular}%
    }
  \label{tab:ds_syn}%
\end{table}%

\subsubsection{Real-world Datasets}
Three real-world surgical instrument segmentation datasets, i.e., Endo17~\cite{allan20192017}, Endo18~\cite{allan20202018}, and RoboTool~\cite{garcia2021image}, are utilized in our experiments. 
% , as shown in Fig.~\ref{fig:src_imgs}
% There are eight types of instruments in the Endo18~\cite{allan20202018} dataset, i.e., Fenestrated Bipolar Forceps, Maryland Bipolar Forceps, Prograsp Forceps, Large Needle Driver, Monopolar Curved Scissors, Ultrasound Probe, Suction Instrument, and Clip Applier.
% The RoboTool~\cite{garcia2021image} dataset covers 20 different surgical procedures.
We train on the training split of the Endo18~\cite{allan20202018} dataset with 2235 images to obtain an upper bound of the binary segmentation task. 
% with all the training-time augmentations applied.
The test split of Endo18~\cite{allan20202018} dataset with 997 images is used to evaluate the in-distribution performance, while the Endo17~\cite{allan20192017} dataset with 1800 images and the RoboTool~\cite{garcia2021image} dataset with 514 images are used for cross-domain out-of-distribution evaluation. Besides the visually distinctive background tissues difference, novel types of instruments like the Vessel Sealer and Grasping Retractor in Endo17~\cite{allan20192017} dataset and new instruments-tissue interactions in RoboTool~\cite{garcia2021image} dataset pose increasing challenges to the model to generalize and maintain acceptable performance. The RoboTool~\cite{garcia2021image} dataset has a much larger domain gap than Endo17~\cite{allan20192017} dataset compared with the Endo18~\cite{allan20202018} dataset, causing over 30\% degradation of upper bound performance, as shown in Table~\ref{tab:overall}.

\begin{figure*}[!htbp]
\centering
\includegraphics[width=\textwidth]{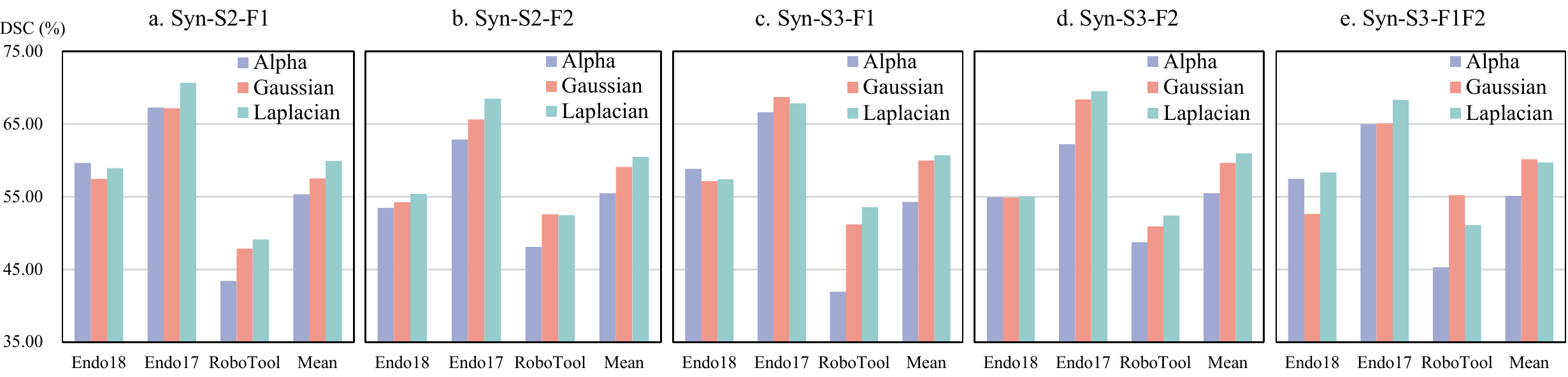}
\caption{\textbf{Comparison between three blending modes under different synthetic dataset settings.} The DSC results on each test dataset and the average results of Endo17~\cite{allan20192017} and RoboTool~\cite{garcia2021image} are reported.
In most cases, Laplacian Blending yields optimal performance.
}
\label{fig:blend_mode}
\end{figure*}

\subsection{Implementation Details}
We utilize the classical UNet~\cite{ronneberger2015u} architecture to implement our binary segmentation model for all the experiments. Binary cross-entropy loss and the Adam~\cite{kingma2014adam} optimizer 
% from the PyTorch~\footnote{\url{https://pytorch.org}} library 
are adopted to train the model with a batch size of 64 and a learning rate of 0.001 on the NVIDIA RTX3090 GPU platform for 100 epochs. The images are resized to $224\times224$ in height and width.
% to save training time and memory consumption. 
We use the Dice Similarity Coefficient (DSC) as the evaluation metric to compare the segmentation performance. 
% The three blending techniques, i.e., 
Alpha Blending, Gaussian Blending, and Laplacian Blending are implemented referring to the semi-synthetic~\footnote{https://github.com/luiscarlosgph/semi-synthetic} repository.
% from Garcia-Peraza-Herrera et al.~\cite{garcia2021image}. 
Besides, 
% three training-time augmentations in our pipeline, i.e., the 
Chained Augmentation Mixing (CAM), Coarsely Dropout (CDO), and Element-wise Patch Mixing (EPM) are adapted from the official Augmix~\cite{hendrycks2019augmix} repository, 
% repository~\footnote{\url{https://github.com/google-research/augmix}}, 
the albumentation~\footnote{https://github.com/albumentations-team/albumentations} library, and the timm~\footnote{https://github.com/rwightman/pytorch-image-models} library with their default parameters unless otherwise specified. The foreground and background seed images and detailed codes of implementation and adaptation are available at~\url{https://github.com/lofrienger/OneToMany_ToolSynSeg}.

\section{Results and Analysis}
\label{sec:res}

We evaluate our framework with multiple real-world datasets, i.e., the Endo18~\cite{allan20202018} test dataset, the Endo17~\cite{allan20192017} dataset, and the RoboTool~\cite{garcia2021image} dataset. The performance of the real dataset Endo18~\cite{allan20202018} is taken as the in-distribution baseline and upper bound for comparison. Laplacian Blending is utilized as the blending operation and the proposed hybrid augmentation is applied during training. The overall results are shown in Table~\ref{tab:overall}. Our synthetic datasets can achieve decent results on all test datasets. Especially on the RoboTool~\cite{garcia2021image} dataset, our best result outperforms the baseline with 8.35\% DSC. 
On average of two out-of-distribution datasets, i.e., Endo17~\cite{allan20192017} and RoboTool~\cite{garcia2021image}, our approach yields 0.76\% DSC gain, showing outstanding generalization performance. Compared with Endo17~\cite{allan20192017}, the RoboTool~\cite{garcia2021image} dataset contains more complex surgical scenes including various instrument actions and tissue-tool interactions, making it more challenging for the model to generalize well. As indicated in Table~\ref{tab:overall}, the baseline model suffers over 30\% DSC drop compared with its performance on the test split of Endo18~\cite{allan20202018}. Whereas, the performance degradation of our method is much smaller and acceptable. Hence, our approach, based on synthetic data, demonstrates superiority in addressing domain shifts and complex scene segmentation challenges.

% All augmentations 
\begin{table}[!htbp]
  \centering
  \caption{\textbf{Overall quantitative results of our synthetic datasets.}
  The best and runner-up results from our datasets are indicated in bold and underlined. 
  }
    \resizebox{\columnwidth}{!}{%
    \begin{tabular}{ccccc}
    \toprule
    \multirow{1}[4]{*}{Training} & \multicolumn{4}{c}{Test DSC (\%)} \\
\cmidrule{2-5}          & Endo18~\cite{allan20202018} & Endo17~\cite{allan20192017} & RoboTool~\cite{garcia2021image} &   Mean$\pm$STD \\
    \midrule
    Endo18~\cite{allan20202018} & 85.04 & 86.78 & 52.44 & 69.61 $\pm$ 17.17 \\
    \midrule
    Syn-S2-F1  & 71.65 & 76.61 & 55.01 & 65.81 $\pm$ 10.80 \\
    Syn-S2-F2  & 71.75 & \underline{78.47} & 56.75 & 67.61 $\pm$ 10.86 \\
    Syn-S3-F1 & \underline{71.84} & 78.18 & 59.45 & \underline{68.82 $\pm$ 9.37} \\
    Syn-S3-F2 & \textbf{72.07} & 76.08 & \underline{60.68} & 68.38 $\pm$ 7.70 \\
    Syn-S3-F1F2  & 71.55 & \textbf{79.95} & \textbf{60.79} & \textbf{70.37 $\pm$ 9.58} \\
    \bottomrule
    \end{tabular}%
    }
  \label{tab:overall}%
\end{table}%

\begin{figure}[ht]
\centering
\includegraphics[width=\columnwidth]{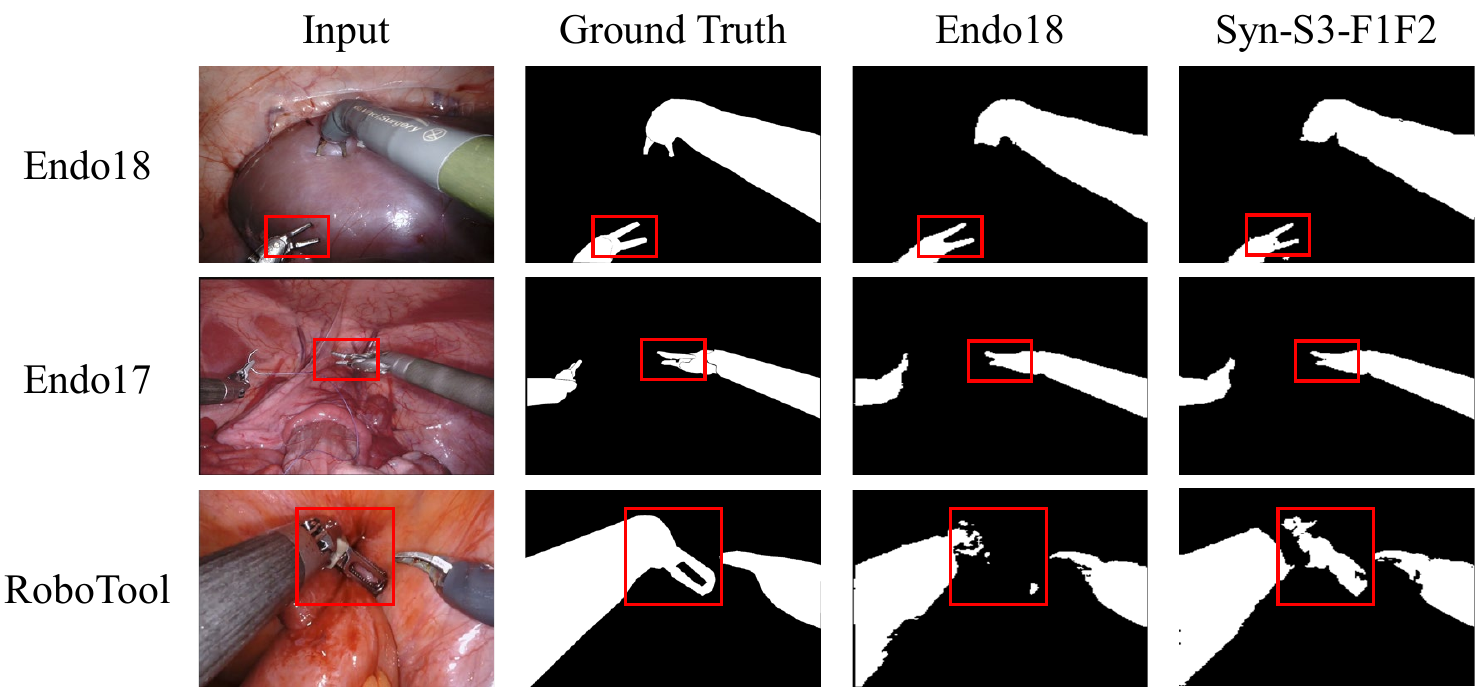}
\caption{\textbf{Qualitative comparison between models trained with the real and synthetic datasets.} Our method can produce comparable and even better segmentation results.
}
\label{fig:quali_cmp}
\end{figure}

% Qualitative Comparison.========================
As shown in Fig.~\ref{fig:quali_cmp}, compared with the model derived from the real Endo18~\cite{allan20202018} dataset, the model trained with our Syn-S3-F1F2 dataset yields competitive segmentation performance on the test images from three real datasets. Notably, as highlighted in the rectangle regions, the clasper parts of the instruments are well recognized, especially for the image from RoboTool~\cite{garcia2021image}.

% Table generated by Excel2LaTeX from sheet 'main'
\begin{table*}[ht]
  \centering
  \caption{\textbf{Ablation study about the hybrid training-time augmentation.} CAM, CDO, and EPM are short for Chained Augmentation Mixing, Coarsely Dropout, and Element-wise Patch Mixing, respectively. Results show the efficacy of the proposed hybrid training-time augmentation.
%   These three augmentations all help improve the segmentation results, and their combination provides the optimal performance on average. 
  The best and runner-up results are in bold and underlined.}
  \label{tab:aug_ablation}
    \resizebox{0.7\textwidth}{!}{%
    % \resizebox{\columnwidth}{!}{%
    \begin{tabular}{cccccccc}
    \toprule
    \multirow{1}[4]{*}{Training} & \multicolumn{3}{c}{Augmentation} & \multicolumn{4}{c}{Test DSC (\%)} \\
\cmidrule(lr){2-4} \cmidrule(l){5-8} & CAM & CDO & EPM & Endo18~\cite{allan20202018} & Endo17~\cite{allan20192017} & RoboTool~\cite{garcia2021image} & Mean $\pm$ STD \\
    \midrule
    Endo18~\cite{allan20202018} & \ding{51}     & \ding{51}     & \ding{51}     & 85.04 & 86.78 & 52.44 & 69.61$\pm$17.17 \\
    \midrule
    \multirow{5}[2]{*}{Syn-S2-F1} & \ding{55}     & \ding{55}     & \ding{55}     & 59.67 & 67.26 & 43.45 & 55.36 $\pm$ 11.91 \\
          & \ding{51}     & \ding{55}     & \ding{55}     & 70.31 & 73.70 & 49.20 & 61.45 $\pm$ 12.25 \\
          & \ding{51}     & \ding{51}     & \ding{55}     & 69.81 & 74.36 & \textbf{58.04} & \textbf{66.20 $\pm$ 8.16} \\
          & \ding{51}     & \ding{55}     & \ding{51}     & \underline{71.35} & \textbf{77.15} & 50.22 & 63.69 $\pm$ 13.47 \\
          & \ding{51}     & \ding{51}     & \ding{51}     & \textbf{71.65} & \underline{76.61} & \underline{55.01} & \underline{65.81 $\pm$ 10.80} \\
    \midrule
    \multirow{5}[2]{*}{Syn-S2-F2} & \ding{55}     & \ding{55}     & \ding{55}     & 55.39 & 68.52 & 52.45 & 60.49 $\pm$ 8.04 \\
          & \ding{51}     & \ding{55}     & \ding{55}     & 71.42 & 75.47 & 54.06 & 64.77 $\pm$ 10.71 \\
          & \ding{51}     & \ding{51}     & \ding{55}     & 70.70 & 76.82 & \underline{54.63} & \underline{65.73 $\pm$ 11.10} \\
          & \ding{51}     & \ding{55}     & \ding{51}     & \textbf{71.93} & \underline{76.99} & 53.33 & 65.16 $\pm$ 11.83 \\
          & \ding{51}     & \ding{51}     & \ding{51}     & \underline{71.75} & \textbf{78.47} & \textbf{56.75} & \textbf{67.61 $\pm$ 10.86} \\
    \midrule
    \multirow{5}[2]{*}{Syn-S3-F1} & \ding{55}     & \ding{55}     & \ding{55}     & 58.86 & 66.65 & 41.92 & 54.29 $\pm$ 12.37 \\
          & \ding{51}     & \ding{55}     & \ding{55}     & 70.05 & 74.54 & 49.97 & 62.26 $\pm$ 12.29 \\
          & \ding{51}     & \ding{51}     & \ding{55}     & 69.69 & 75.04 & \underline{53.88} & 64.46 $\pm$ 10.58 \\
          & \ding{51}     & \ding{55}     & \ding{51}     & \textbf{72.28} & \textbf{78.21} & 52.33 & \underline{65.27 $\pm$ 12.94} \\
          & \ding{51}     & \ding{51}     & \ding{51}     & \underline{71.84} & \underline{78.18} & \textbf{59.45} & \textbf{68.82 $\pm$ 9.37} \\
    \midrule
    \multirow{5}[2]{*}{Syn-S3-F2} & \ding{55}     & \ding{55}     & \ding{55}     & 55.05 & 69.52 & 52.43 & 60.98 $\pm$ 8.55 \\
          & \ding{51}     & \ding{55}     & \ding{55}     & 70.96 & \underline{77.33} & 55.11 & 66.22 $\pm$ 11.11 \\
          & \ding{51}     & \ding{51}     & \ding{55}     & \underline{72.02} & \textbf{77.39} & \underline{58.34} & \underline{67.87 $\pm$ 9.52} \\
          & \ding{51}     & \ding{55}     & \ding{51}     & 71.68 & 75.48 & 54.36 & 64.92 $\pm$ 10.56 \\
          & \ding{51}     & \ding{51}     & \ding{51}     & \textbf{72.07} & 76.08 & \textbf{60.68} & \textbf{68.38 $\pm$ 7.70} \\
    \midrule
    \multirow{5}[2]{*}{Syn-S3-F1F2} & \ding{55}     & \ding{55}     & \ding{55}     & 58.33 & 68.34 & 51.10 & 59.72 $\pm$ 8.62 \\
          & \ding{51}     & \ding{55}     & \ding{55}     & 71.46 & 75.97 & 53.52 & 64.75 $\pm$ 11.23 \\
          & \ding{51}     & \ding{51}     & \ding{55}     & 71.11 & 76.85 & \underline{57.71} & \underline{67.28 $\pm$ 9.57} \\
          & \ding{51}     & \ding{55}     & \ding{51}     & \textbf{71.97} & \underline{78.30} & 54.18 & 66.24 $\pm$ 12.06 \\
          & \ding{51}     & \ding{51}     & \ding{51}     & \underline{71.55} & \textbf{79.95} & \textbf{60.79} & \textbf{70.37 $\pm$ 9.58} \\
    \bottomrule
    \end{tabular}
     }
\end{table*}%

\subsection{Choices of Blending Mode}
% When composing the sampled foreground instrument and the background tissue images, different blending modes . 
To find the preferable blending mode which produces softer blending artifacts, we compare three blending modes, i.e., Alpha Blending, Gaussian Blending, and Laplacian Blending. As shown in Fig.~\ref{fig:blend_mode}, under five synthetic dataset settings, Laplacian Blending yields the optimal performance for most test datasets. By constructing the Laplacian pyramid for both the foreground and background images, the blending process can take into account the details and structures at different scales. As a result, the boundaries between the foreground and background images are smoother and more seamless, resulting in a more realistic and visually pleasing composite image. 
% Hence, Laplacian Blending is the prior choice for image composition-based surgical scene synthesizing. 

\subsection{Ablation Analysis of Hybrid Training-time Augmentation}
We further conduct the ablation study about three advanced training-time augmentations in our pipeline, i.e., Chained Augmentation Mixing (CAM), Coarsely Dropout (CDO), and Element-wise Patch Mixing (EPM). As shown in Table~\ref{tab:aug_ablation}, 
% for all the five synthetic training datasets,
% when training with our mixed synthetic dataset, 
these training-time augmentations steadily help improve the segmentation performance on the real test datasets. In most cases, their hybrid integration yields the best average results on two out-of-distribution datasets, i.e., Endo17~\cite{allan20192017} and RoboTool~\cite{garcia2021image} and provides 7.12\% to 14.53\% DSC gain compared to not having it, reflecting its great benefit in boosting model generalization capability.

% Table generated by Excel2LaTeX from sheet 'joint'
\begin{table}[!htbp]
  \centering
  \caption{\textbf{Segmentation results of semi-supervised synthetic-real joint training.} 
  A small ratio of real data can yield significant performance gains when jointly trained with our synthetic dataset. 
The best and runner-up results are in bold and underlined, respectively.}
    \resizebox{\columnwidth}{!}{%
    \begin{tabular}{cccccc}
    \toprule
    \multicolumn{2}{c}{Training} & \multicolumn{4}{c}{Test DSC (\%)} \\
\cmidrule(r){1-2} \cmidrule(l){3-6}    Data & Size & Endo18~\cite{allan20202018} & Endo17~\cite{allan20192017} & RoboTool~\cite{garcia2021image} & Mean $\pm$ STD \\
    \midrule
    \multicolumn{1}{c}{Syn-S3-F1F2} & 2235 & 71.55 & 79.95 & 60.79 & 70.37 $\pm$ 9.58 \\
    \midrule
    \begin{tabular}[c]{@{}c@{}}90\% Syn-S3-F1F2 \\ + 10\% Endo18~\cite{allan20202018}\end{tabular} & 2235 & 76.86 & 81.91 & 61.63 & 71.77 $\pm$ 10.14 \\
    % \cmidrule(r){1-2}
    \begin{tabular}[c]{@{}c@{}}80\% Syn-S3-F1F2 \\ + 20\% Endo18~\cite{allan20202018}\end{tabular} & 2235 & 79.91 & 83.35 & 61.05 & 72.20 $\pm$ 11.15 \\
    \midrule
    \begin{tabular}[c]{@{}c@{}} Syn-S3-F1F2 \\ + 10\% Endo18~\cite{allan20202018}\end{tabular} & \begin{tabular}[c]{@{}c@{}} 2235 \\ + 223\end{tabular} & 79.32 & 84.32 & \underline{65.57} & \underline{74.95 $\pm$ 9.38} \\
    % \cmidrule(r){1-2}
    \begin{tabular}[c]{@{}c@{}} Syn-S3-F1F2 \\ + 20\% Endo18~\cite{allan20202018}\end{tabular} & \begin{tabular}[c]{@{}c@{}} 2235 \\ + 447\end{tabular} & \underline{81.66} & \underline{85.91} & \textbf{65.92} & \textbf{75.92 $\pm$ 10.00} \\
    \midrule
    \multicolumn{1}{c}{Endo18~\cite{allan20202018}} & 2235 & \textbf{85.04} & \textbf{86.78} & 52.44 & 69.61 $\pm$ 17.17 \\
    \bottomrule
    \end{tabular}%
    }
  \label{tab:joint_tr}%
\end{table}%

\subsection{Semi-supervised Synthetic-real Joint Training}
Although abundant well-annotated data are expensive and scarce, a small ratio of them is typically affordable for preliminary investigation. 
% Considering this, 
To explore the broader impact of our framework, we conduct experiments following the scheme of semi-supervised synthetic-real joint training, where a small portion of real data is utilized for training jointly with the generated synthetic dataset.
% and the results suggest its great benefit in boosting performance. 
The results of pure Syn-S3-F1F2 and pure Endo18~\cite{allan20202018} are treated as two references. As shown in Table~\ref{tab:joint_tr}, when keeping the same training sample size and replacing 20\% synthetic data with the real data, the average DSC of two out-of-distribution domains, i.e., Endo17~\cite{allan20192017} and RoboTool~\cite{garcia2021image} gets improved with 2.59\% compared with the reference result of Endo18~\cite{allan20202018}. 
Further, when adding 20\% real data on top of the entire pure Syn-S3-F1F2 dataset, on average, the performance gain is increased by 6.31\% DSC, indicating superior generalization ability. The performance on the in-distribution domain Endo18~\cite{allan20202018} also gets boosted with 10.11\% DSC compared with the reference result of the pure Syn-S3-F1F2 dataset.
% The much lower standard deviations in average results also indicate the stability and reliability of such a combined data training scheme.
% Hence, with greatly reduced effort of data preparation,  is of practical benefit to provide superior generalization performance. 

\section{Conclusion and Future Work}
This work proposes a highly efficient one-to-many data-centric framework for surgical instrument synthesis and segmentation. Instead of model architecture design and optimization, we focus on improving the quality of synthetic data under extreme source constraints to alleviate the challenging data-related issues in surgical instrument segmentation, i.e., data shortage and shift.

Specifically, for data synthesizing, we only leverage one background tissue image and a few foreground instrument images to construct the source image pools with varieties of augmentations. Then one tissue image and one or two instrument images are randomly sampled from the background and foreground image pools and blended together to form the final synthetic surgical scene image. We extensively explore and compare three blending techniques, i.e.,  Alpha Blending, Gaussian Blending, and Laplacian Blending, and find Laplacian Blending to be the optimal choice. With the generated synthetic datasets, we incorporate the hybrid usage of advanced training-time augmentations, i.e., Chained Augmentation Mixing (CAM), Coarsely Dropout (CDO), and Element-wise Patch Mixing (EPM) when training the segmentation model. The generalization ability across multiple real test datasets gets steadily improved with the proposed hybrid augmentation. Moreover, we also demonstrate the efficacy of our framework in the semi-supervised synthetic-real joint training scheme to help boost the in-distribution and out-of-distribution performance.

\textbf{Future work} can extend the proposed framework in two major directions. On the one hand, 
% more approaches, such as creating new instrument viewpoints with Nerf~\cite{mildenhall2021nerf}, can be explored to further improve the quality of the synthetic data. Besides, 
more elements in real surgical scenes, like blood and smoke, can also be synthesized to increase the overall realism of the dataset. On the other hand, domain adaptation methods can also be introduced when training the segmentation model to mitigate the discrepancy between the synthetic and real domains. We expect that our approach can encourage more data-efficient and data-driven approaches in surgical scene understanding applications when dealing with data-related issues, such as data shortage, domain shift, class imbalances, and incremental classes, to name a few.

% \backmatter

% \bmhead{Supplementary information}
% The online version contains the mentioned supplementary material.
% If your article has accompanying supplementary file/s please state so here. 

% Authors reporting data from electrophoretic gels and blots should supply the full unprocessed scans for key as part of their Supplementary information. This may be requested by the editorial team/s if it is missing.

% Please refer to Journal-level guidance for any specific requirements.

% \bibliographystyle{ieeetr}
\bibliographystyle{plain}
\bibliography{mybib}% common bib file



\end{document}